# EPIDEMIOLOGY THROUGH CELLULAR AUTOMATA
## Case of Study: Avian Influenza in Indonesia [*]

Hokky Situngkir
(hokky@elka.ee.itb.ac.id)
Dept. Computational Sociology
Bandung Fe Institute

## Abstract

This paper performs the utilization of cellular automata computational analysis as the dynamic model of spatial epidemiology. Here, explored elementary aspects of cellular automata and its application in analyzing contagious disease, in this case avian influenza disease in Indonesia. Computational model is built and map-based simulation is performed using several simplified data of such transportation through sea in Indonesia, and its accordance with poultries in Indonesia, with initial condition of notified avian influenza infected area in Indonesia. The initial places are Pekalongan, West Java, East Java, and several regions in Sumatera. The result of simulation is showing the spreading-rate of influenza and in simple way and describing possible preventive action through isolation of infected areas as a major step of preventing pandemic.

Key Words: epidemiology, cellular automata, spatial model, avian influenza, Indonesia

---

*Epidemics have often been more influential than statesmen and soldiers in shaping the course of political history, and diseases may also color the moods of civilizations.*
The White Plague - René Dubos (1901 - 1982)

## 1. Background

As an agricultural country, the beginning of the year 2004 gave a surprise on death of millions of chickens throughout the country – at first, the disease was regarded as common poultry disease, yet in fact the disease is a result of the activity of a kind of mutant influenza virus that even deadly to human. Ever since November 2003, there have been 5 millions of chickens died as the result of this virus activity in Indonesia. There are few kinds of avian influenza (*avian influenza*), which first detected in Hong Kong in the year 1997 – it was reported 18 citizens positively infected with 6 of them died. In virology, this virus was reviewed in Hatta & Kawaoko (2002).

Before the issue spreads in Indonesia, the disease that primarily infecting chicken has been widely a hot gossip in the opening January 2004 in some neighbor countries i.e.: Thailand and Vietnam. As noted by the World Health Organization WHO (2004), the virus attacks human through poultry animals – especially human in the area of poultry aviary. There has not been case showing virus transmission from human to human.

In a glance, this epidemic may be looked simple, since human transmission could only occur through intensive interaction with infected poultry. Furthermore, this influenza virus cannot transmit to human through eating the cooked since the cooked one will soon be penalized by certain temperature – like the other influenza virus. Yet, of course, this human infection limitation cannot be ignored since the large number of Indonesian employees working, even live in the poultry areas (chickens, ducks, etc.).

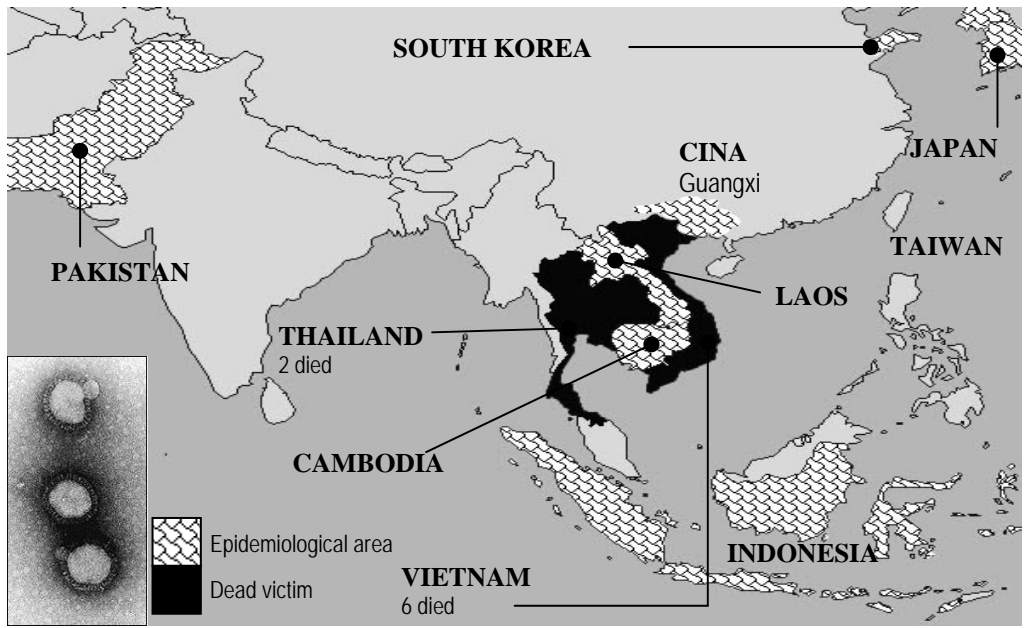

**Figure 1**
The spreading of Avian influenza in Southeast Asia Region up to January 27, 2004 according to WHO. *Inset:* The Avian Influenza virus - *CDC(2004)*

The paper aims to give description on how epidemiology has also been enriched with one of computational complexity tools, the cellular automata – as a spatial and discrete model of dynamical system - in this case the spreading of contagious disease. The paper will not discuss virology and biological aspects in details, yet it elaborates more on epidemiology. There are four major sections in the paper. The first one describes the epidemiological aspects of a contagious disease, which can be portrayed using cellular automata analysis.

The second describes the spreading of avian influenza through mathematical and computational model. The third section of the paper discusses the result of the simulation performed. The paper is ended by drawing some concluding remarks and few directions of preventing avian influenza pandemics as a result of the simulation performed.

## 2. Epidemic Automata

Epidemiology enriched by cellular automata analysis is an interdisciplinary synthetic approach between bio-mathematic and dynamic computational simulation. Interdisciplinary approach is urgently needed to understand the spreading of a disease in a social system. As explained by Angulo (1997) that interdisciplinary approach on epidemics basically should acquire analytical tools in biology, social-behavioral, geography in spatial approach, mathematics and computation, even economy and cultural analysis. This awareness has brought us into endeavor on differing the two approaches that can bring us to the synergetic of biomathematics and computational sociology approach.

### 2.1. A Short Introduction to Epidemiological Model

Epidemiology is a discourse of the spreading disease or in deeper meaning the spreading of elementary foundation of social system. The spreading disease in certain area or locally is called as endemic, and if it spreads to several level of endemics should be called as epidemic. The large (world) scale of epidemic is called as pandemic. Epidemiology is an integral approach over nature, social sanitation, economy, even military, which makes epidemiology discussion become a very extensive discussion. In other words, modeling in epidemiology needs many approaches over living organism ecology holistically. In major, epidemiology discusses how living organisms can survive within their ecosystem.

Nowadays epidemiology has known numerous disease-spreading models. One of the famous models is the stochastic model, such as model of Kermack and McKendrick (Bartholomew, 1982:248-272). This elaboration result is then verified with the result of our experimental simulation within the second section of the paper. The usual innovation epidemiological model distinguishes the susceptibility of automaton to be approached by a disease, get infected, and then recover or died – this model is known with SIRS model (computational elaboration for this model can be reviewed in Piyawong, *et.al.*, (2003). Epidemiology model using cellular automata is a model that focuses on spatial spreading of a disease. Means, this perspective tries to grasp structure of the spreading occurs to be then simulated computationally.

### 2.2. Cellular Automata on Epidemics

At the beginning, cellular automata were a model in physics to analyze the dynamics of micro-scale particle, such as thermodynamic rule, *spin-glass* model, etc (Wolfram, 1983). Yet in general, we may say that cellular automata is dynamical system that is formed base upon spatial and discrete time as model of physical process as computational device (Mitchell, *et.al.*, 1990). Cellular automata consist of spatial grids placed by cells in certain time (t) and state. At each discrete time, we perform iteration in which cells update on certain rules. In this case we can say that cellular automata are a perfect feedback machine, or more specific, a finite state machine, which its state changes step by step.

Dimensionally, we can categorize cellular automata to be one, two, or three dimensions. This categorization is surely easier to understand by looking at how the pattern of cellular automata run or iterated. In epidemic model, we ground ideas to the reality of spatial system. One of the utilization of cellular automata 2-D is described in Figure 2: the movement of pseudopodia of amoeba. Cellular automata 2-D has been known in biometry study and organism behavioural study – the most famous application is the artificial life application, which explains behavioural structure of organism spatially. The detail introduction to cellular automata model can be seen in Chopard & Droz (1998).

In the Cellular automata we know two major concepts, i.e.:
- The rules of micro dynamics of the cell's state
- Lattice (*grid*) to represent the spatial positions of each automata.

These two concepts give picture on the dynamic of each automata cell represents object we observe. As a dynamic model for epidemiology, we represent the disease spreading using epidemiological parameters into cellular automata rules and how it spreads within automata positional unit.

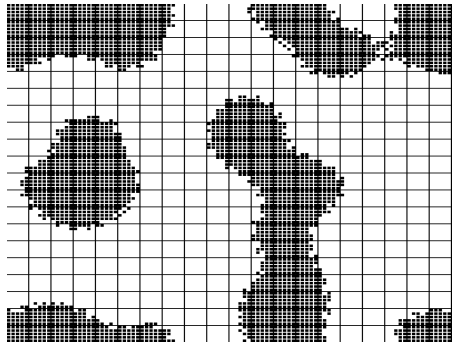
**Figure 2**
Imitation amoeba movement (single cell animal) with their pseudopodia movement using cellular automata 2-D.

Noticing simple model of spreading disease as explained above, in epidemiology at least we recognize 3 states of automata, i.e.:
1. *Susceptible* condition that is a condition where a population agent has not yet infected, but has certain probabilistic potential to be infected by the disease.
2. *Infected* condition that is a condition when an agent infected.
3. *Recovery* condition that is a condition when the disease disappears from the agent; be it recover or die.

In epidemiology dynamics, we also represent population computational parameters, they are:
- Neighborhood relation (interconnectedness of one agent with other agent) spatially or in network (agent spatially does not stay aside, but it has close inter-relation, such as transportation, etc).
- Probability for someone to be infected and one's capability to recover or die from the disease.
- Phases as the result of disease infection and probability of recover person and to be re-infected.

From this point we analyze how cellular automata model the epidemiology of avian influenza disease through slight review of ecological pattern of its spreading.

## 3. Avian influenza Epidemiology Pattern

Avian influenza is a result of a very unique virus activity. It spreads from poultry to other poultries with a high mutation rate. It attacks poultry, and may frantically attack other species of animal e.g.: horse, pig, sea mammals, even, to human (Brown, 2001 and Capua & Alexander, 2002). The epidemiological pattern is also unique. Until the paper is written, there has not been any report of infection through inter-human contact, nor infection to human through infected poultry food consumption. It attacks human through direct contact with living infected poultry or animals. As elaborated in Capua & Alexander (2002), major means of infection are their saliva, feces, and other object like cages, etc. From this we may conclude that the very risky places in the case of avian influenza disease epidemiology are:
- Poultry areas
- Trading center of living poultry
- Transportation area of living poultry distribution

The interesting part of this specific epidemiology is that it contains two stages of spreading, as illustrated in Figure 3. In the picture we can see that the first stage is the virus spreading between contact and interacting poultry, and between infected poultry with other animals such as pigs or horses, etc. Human life survival is threatened by interaction between human and the infected animals. Here we detect 3 epidemic coefficients, they are $\alpha$, as the coefficient that depends on virus transportation through distributed poultry, $\beta$, epidemic coefficient, which concerns with inter-poultry animals contact in poultry area, and $\gamma$, that is epidemic coefficient which depends on human interaction with the infected animals. From these three interactions, we can say that probabilistic relation from each variable follows:

$$\alpha < \beta > \gamma \ ^1 \qquad \ldots(1)$$

Formally, can be written down as:

$$\frac{dS_2}{dt} \approx \alpha S_1 + \beta S_2 \qquad \ldots(2)$$

$$\frac{dI}{dt} \approx \gamma S_2 - mI \qquad \ldots(3)$$

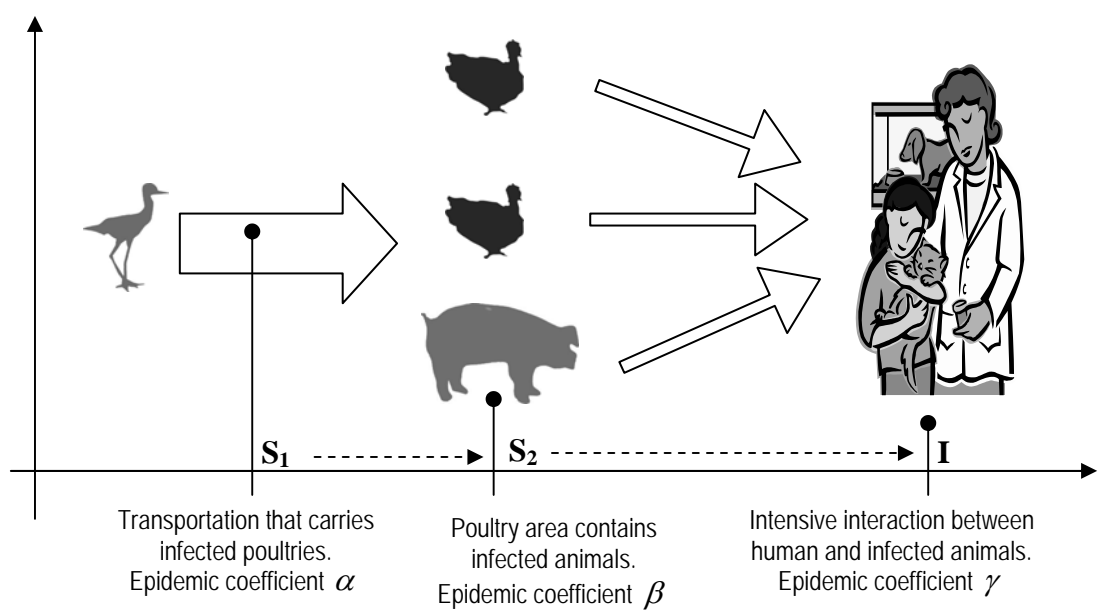

**Figure 3**
The spreading pattern of bird flu disease

This formalism is not based on Kermack and McKendrick's (Willox, *et.al.*, 2003) epidemiology equations by eliminating *recovery* factor and $m$, death factor caused by the flu infection. As the consequence, we can say that this model is focusing in virus diffusion occurs in society. The epidemic model with cellular automata as explained in Rhodes & Anderson (1998) basically inspires this model.

To give spatial description, we performed discrete Indonesia territorial map by grids. Based on the data of poultry area issued by the government, Dirjen Bina Produksi Peternakan (2003), basically we can see that each area in those grids owns folk poultry area that possibly infected and also living poultry market throughout the archipelago – this is assumed for the sake of simplicity. On contact probability between islands, we use naval map PELNI (Loud, 2003) showing major port locations in Indonesia. From here we can pick some major ports that are possibly used as poultry transportation route.

On the terrestrial interaction, we use von Neumann neighborhood system, as illustrated in figure 4, that the probability of epidemic occurs in an area with its neighbor area in the north, south, wet, and east side. For possible infected human, we give certain probability by seeing if the neighbor area thoroughly infected or not.

---

[1] This relation is made by facts that animals in one poultry area are easier to infect each other relative to infection in human and by transportation of infected animals.

By picking $\alpha, \beta, \gamma$ value according to (1), we develop computational algorithm to simulate the diffusion of avian influenza virus in Indonesia, in major outline as follow:

```
begin
      spatial map initiation;
      build network_S2
            major_ports := tg_priok ⊕ merak ⊕ bakauheni...
            ⊕ pontianak&kepatang ⊕ balikpapan&samarinda...
            ⊕ makasar&parepare ⊕ ... etc;
      terrestrial_interaction := von_neuman neighborhood;
      infection_initial_condition;

for i := 1 up to number_iteration
      begin
            interaction_S1 with probability alpha;
            interaction _S2 with probability beta;
            interaction _I; with probability gamma;
      end

end
```

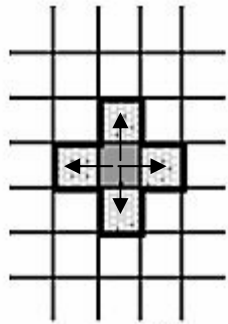

**Figure 4**
von Neumann - Neighborhood scheme used in simulation

## 5. Results & Analysis

We run the simulation by giving initial condition of infected areas in Pekalongan (Central Java), East Java, West Java, and Sumatra with important coefficient values as illustrated in the previous section. The result we gain is illustrated in Figure 5.

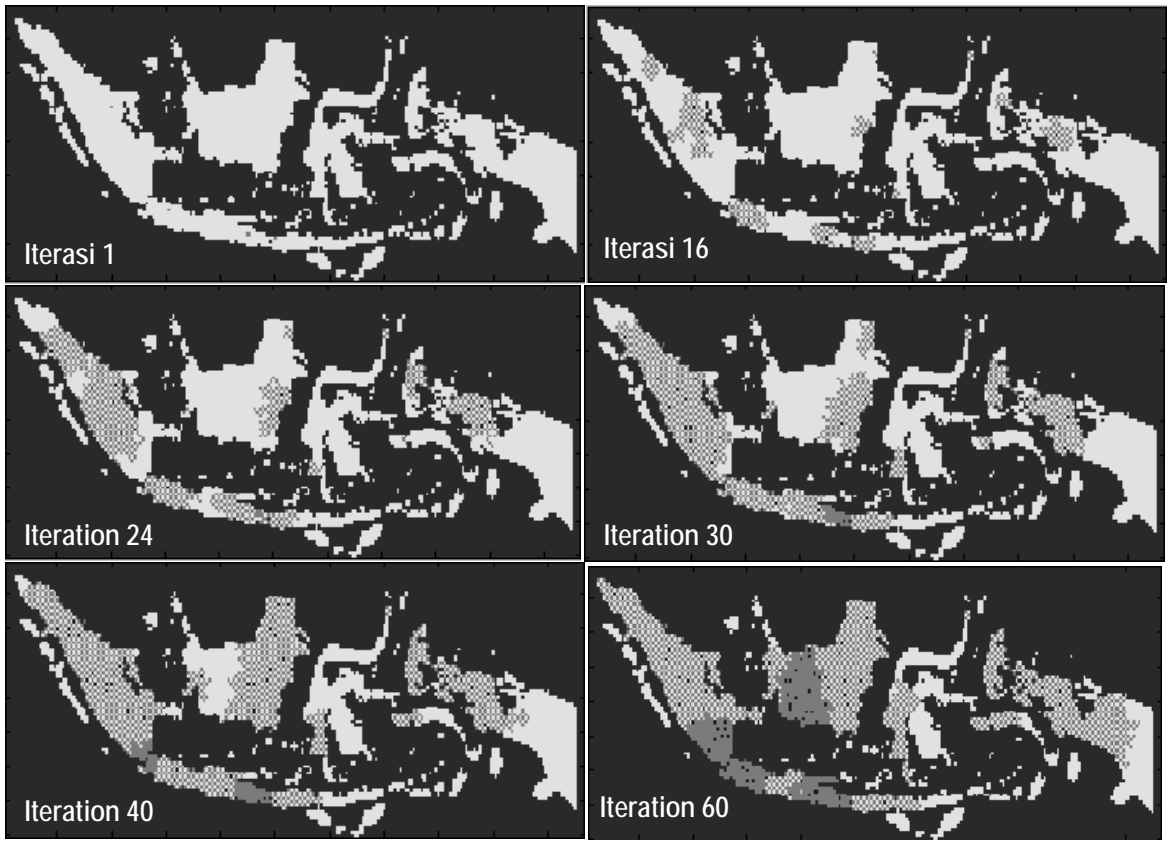

**Figure 5**
Discrete-spatial simulation result performed. Grey spots assign infected areas ($S_2$) and black spots assign high probability for infection to human (I).

From the picture we can clearly see how rapid the spreading of avian influenza disease in Indonesia, though the majority area in Indonesia is sea – this is because in almost every spot of islands lies active ports that possibly used for the purpose of infected poultry transportation. To be clearer, the simulation result can be seen in Figure 6.

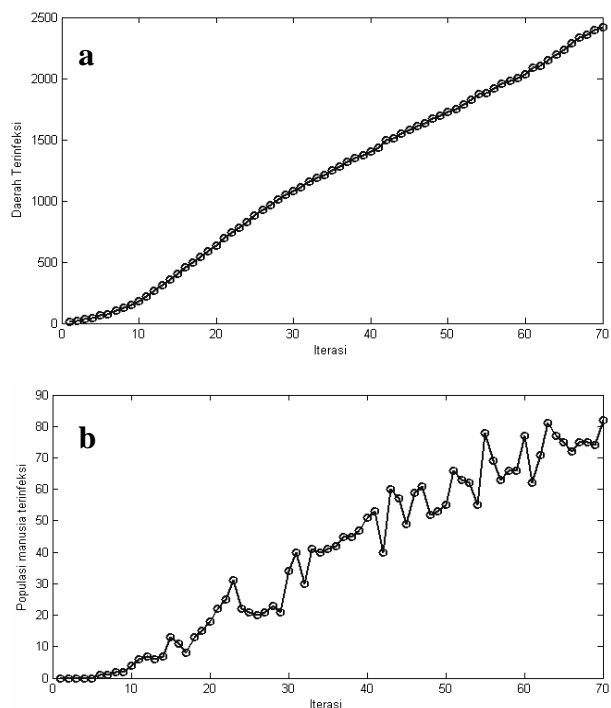

**Figure 6**
The rise of infected areas (a) and number of people infected (b)

On the picture, we can see a monotonic rise from those who got infected by avian influenza; this is because it is assumed that there has not been any significant curative or preventive action in the whole area. The number of people infected is few, yet if there are no serious policies to prevent or at least to slow down the diffusion, the impact will be disastrous since the number of people infected tend to rise. In the other side, number of poultry died in infected areas certainly will cause en economic lost, which in turn will disturb economical system nationwide.

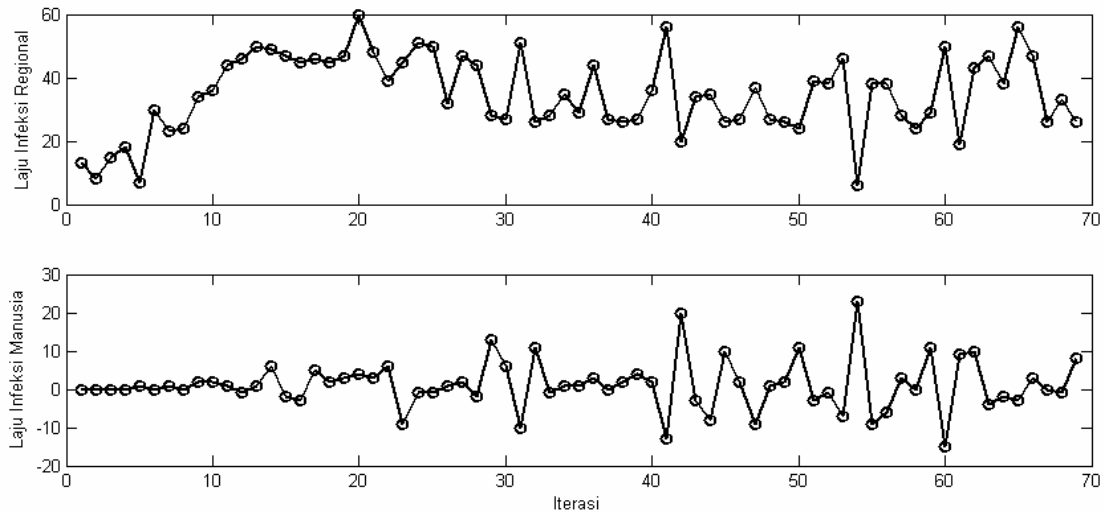

**Figure 7**
The increasing number of areas infected and number of people infected.

Correlating simulation result we obtained with the reality, we can see several hindrances we will find in solving this specific epidemic, i.e.:

- The high rate of diffusion– water area that separates islands does reduce its acceleration, yet it is not significant since the migration of poultry industry is very active.
- The medication for infected human has not been found yet; moreover, the rate of mutation of the virus from time to time is very high. The only medication suggested for the infected people is to having sufficient rest to strengthen self-immunity.
- This problem is highly connected with national economy system that is disrupted by the death of numerous of poultry and export banning from international ally.
- The least serious policy issued by the government obviously affects this epidemic problem. There are many facts on the least considerations on the health of labor forces in Indonesia.

It is important to note that the main goal of the paper is to give description of the diffusion rate of infected area from avian influenza virus, which in fact also brings disaster to human. Before we discuss about medication or other curative action, it is important to reduce the diffusion rate before it becomes pandemic. The main advice suggested to view interdisciplinary complication in the case of avian influenza epidemiology is isolation of detected avian influenza and immediate medical research to have deep knowledge on this virus through medical research, virology, sanitation, etc. Areas that are extremely necessary focus on are poultry areas, market area that provide living poultry, through noticing distribution route of all commodities related to poultry, be it cages, germ/egg, woof, etc. All form of suggestions or warning and *self-help* information about avian influenza apparently cannot reduce the risk of simultaneous diffusion. Infected people or high risk of infection people should be handled in centralistic way by health and poultry instances and other proper instances.

# 7. Concluding Remarks

This paper has shown spatial epidemiology using cellular automata tool. Cellular automata computational tool are giving assistance in understanding the disease spreading by noticing any elements related with the epidemic disease. This shows how important interdisciplinary approach in theoretical construction and constituting policy, in this case, epidemiology.

From the simulation carried out and model constructed, it is obvious that avian influenza epidemic case is an interesting and unique case, regarding to the indirect route of virus to infect human – yet frantically and rapidly attack animals that are close to human and social system where human live. The simulation also showed epidemiology discourse should be put more on priority than curative action since the spreading rate is very high. Isolation of poultry and animal domestication area, monitoring market that trading living animals and distribution of poultry that is high-risk on bringing the virus needs to be major priority. In addition, educational training and *self-help* suggestion should not be the center or focus to repress the virus diffusion.

Avian influenza can disrupt sustainability of human life, moreover the high mutation probability – as this can increase virus ability, its vicious that threaten human life sustainability. Other aspect that also disturbed is economy, regarding the huge number of labor force in national poultry industries.

## Acknowledgements:

The writer would like to thank Prof. Yohanes Surya (Universitas Pelita Harapan) for his support and direction along the research, Dodi Rustandi for providing quantitative sociology and epidemic biology literatures. All false remains the author's.